\documentclass[pre,twocolumn,groupeaddress,showpacs,floatfix]{revtex4}

\usepackage[usenames, dvipsnames]{color}
\usepackage{times}
\usepackage{graphicx}
\usepackage{bm}
\usepackage{amsmath}
\usepackage{amsfonts}
\usepackage{amssymb}
\usepackage{latexsym}
\usepackage{hyperref}
\usepackage{epsf} 
\usepackage[latin1]{inputenc}

\begin{document}

\title{Reply to ``Comment on `Regularizing Capacity of Metabolic Networks' ''}

\author{Carsten \surname{Marr}}
\email{c.marr@jacobs-university.de}
\affiliation{Computational Systems Biology, School of Engineering
  and Science, Jacobs University Bremen, D-28759 Bremen,
Germany}
\author{Mark \surname{M\"uller-Linow}}
\affiliation{Bioinformatics Group, Department of Biology,
  Darmstadt University of Technology, D-64287 Darmstadt, Germany}
\author{Marc-Thorsten \surname{H\"utt} }
\affiliation{Computational Systems Biology, School of Engineering
  and Science, Jacobs University Bremen, D-28759 Bremen,
Germany}

%\date{\today}

\begin{abstract}
  In a recent paper [C.~Marr, M.~M\"uller-Linow, and M.-T.~H\"utt,
  Phys.~Rev.~E \textbf{75}, 041917 (2007)] we discuss the pronounced
  potential of real metabolic network topologies, compared to
  randomized counterparts, to regularize complex binary dynamics. In
  their comment [P.~Holme and M.~Huss, arXiv:0705.4084v1], Holme and
  Huss criticize our approach and repeat our study with more realistic
  dynamics, where stylized reaction kinetics are implemented on sets
  of pairwise reactions. The authors find no dynamic difference
  between the reaction sets recreated from the metabolic networks and
  randomized counterparts. We reproduce the author's observation and
  find that their algorithm leads to a dynamical fragmentation and
  thus eliminates the topological information contained in the
  graphs. Hence, their approach cannot rule out a connection between
  the topology of metabolic networks and the ubiquity of steady
  states.
\end{abstract}


\maketitle

In a recent investigation \cite{marr07_pre} we study transient
dynamics on abstract representations of metabolic networks. Our
binary threshold dynamics is not supposed to model metabolism -- it
rather serves as a dynamic probe, meant to monitor the processing of
perturbations (coming e.g.~from concentration fluctuations) in the
metabolic network. We analyze the emerging binary time series with
entropy-like measures and find a reduced complexity for metabolic
network structures when compared to different null model
architectures. We argue that metabolic network topologies
predominantly dampen perturbations (compared to randomized topologies)
and therefore might contribute to the reliable establishment of steady
states, the major mode of operation of real metabolism.

Contrary to our approach, Holme and Huss \cite{holme07_pp} model
biochemical dynamics with enzyme kinetics. In order to apply the
two-substrate Michaelis-Menten rate law they recreate a set of
pairwise reactions from the substrate graphs of Ma and Zeng
\cite{ma03}. They compare the sets of recreated reactions with
reactions generated from a null model graph with the same degree
distribution and find no difference in the time evolution of the
standard deviation of the metabolites' net fluxes. The authors claim
that their finding contradicts our results, and conjecture the
insignificance of the network structure to the stability of metabolic
steady states \cite{holme07_pp}.

\begin{figure}[b]
\begin{center}
\includegraphics[width=7cm]{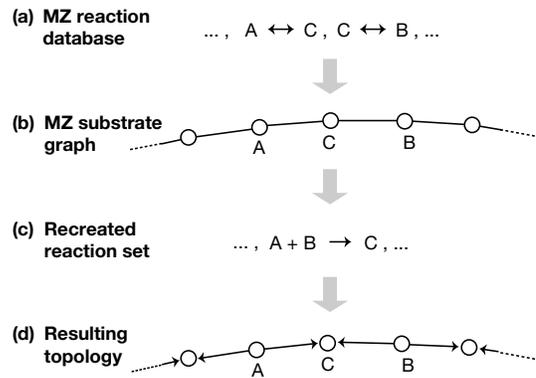}
\caption{The transformation of substrates in the presence
  of current metabolites (a) leads to a linear chains in the metabolic
  networks of Ma and Zeng (b). Via the algorithm in \cite{holme07_pp}, the chain is
  converted into pairwise reactions of the form A + B $\rightarrow$
  C (c). Together with the directionality of the fluxes imposed by the
  reaction kinetics, this leads to a dynamical fragmentation of the
  network. Transport is no longer possible along the resulting
  topology (d).}
\label{chain}
\end{center}
\end{figure}

The approach by Holme and Huss, however, has by construction no access
to the topological information (i.e., larger-scale network properties
beyond pairwise interactions) of interconnected metabolic
reactions. Their algorithm to recreate reaction sets from graphs,
combined with directed kinetics leads to a dynamical fragmentation of
the graph. The information on the topology of metabolism is lost at
this point of the investigation. We want to explain this in detail for
the example of linear chains in the substrate graphs of Ma and Zeng
(MZ) \cite{ma03}, which were used in both \cite{marr07_pre} and
\cite{holme07_pp}. These chains represent the consecutive
transformation of metabolites of the type A $\leftrightarrow$ B. They
are ubiquitous in metabolism and constitute an important topological
feature of the MZ metabolic networks. In most of these reactions,
current metabolites (like ATP, ADP or H$_2$0) act as carriers for
electrons or functional groups, but have been removed from the MZ
reaction database, due to an average path length rationale (see
\cite{ma03} for detailed explanations). The algorithm of Holme and
Huss operates on the undirected MZ graphs assuming that most
generating reactions are of the 2-1-form (A + B $\rightarrow$ C), or
the 2-2-form (A + B $\rightarrow$ C + D). It thus converts a linear
chain of connected metabolites to, at best, topologically connected
reactions with defined directionality, as shown in Fig.~\ref{chain}.
The imposed directionality, however, leads to a dynamical
fragmentation, i.e.~a network of dynamically isolated groups of nodes
where the propagation of information is no longer
possible. Fig.~\ref{chain} illustrates the generation of a chain from
the MZ database and the loss of its information processing potential
during the application of the algorithm of Holme and Huss. This
network 'motif', a directed linear chain, therefore, cannot be
represented with this approach. Note that dynamical fragmentation can
still occur for more highly connected regions of a network. This
effect induces the similarity in dynamics observed in
\cite{holme07_pp} between the recreated metabolic networks and their
randomized counterparts.

\begin{figure}[t]
\begin{center}
\includegraphics[width=7cm]{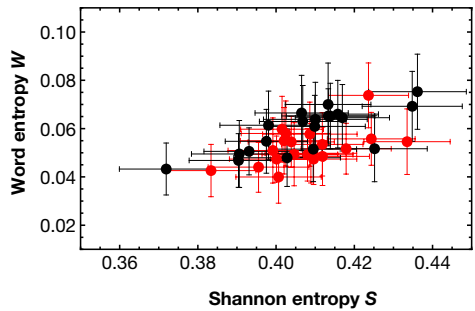}
\caption{The entropy signatures of 40 directed graphs. The entropy
  signatures of 'real' graphs (black) do not separate from those of
  the null model (red) in the entropy plane. All graphs have been
  reconstructed from the reaction sets used in \cite{holme07_pp}, which
  in turn rely on the MZ substrate graph and randomized null models
  of the human metabolic network. Notably, we use
  the same parameters as in \cite{marr07_pre}.}
\label{plane}
\end{center}
\end{figure}

On the grounds of these topological observations, we would expect that
our dynamic probe will also find no difference between the directed
networks reconstructed from the reaction sets from Holme and Huss
and the corresponding null model topologies. The algorithm from
\cite{holme07_pp} provides a protocol to pass from an undirected graph
to a reaction set. We use the standard protocol described in
\cite{ma03} to pass from this reaction set to a graph again. In their
comment, the authors find no dynamical differences between
the 'real' and their null model reaction systems. We implement our
simple binary dynamics on the reconstructed directed graphs and, as
expected, obtain similar entropy signatures for the respective
topologies (see Fig.~\ref{plane}). The low entropy values, compared to
the values of the metabolic networks shown in \cite{marr07_pre} are a
consequence of spatiotemporal patterns of low complexity, which
naturally arise from networks consisting of small fragments, where
global information transport is suppressed.

Many aspects of the relationship between topology and dynamics are
still unclear, especially when it comes to real systems. A fundamental
question raised in \cite{holme07_pp} is, whether the entropy signature
of a network is a meaningful observable or rather a data-mining tool.
We acknowledge this criticism. Up to now, we cannot correlate the
entropy measures with a clear biological quantity. Nevertheless, we
are convinced that mapping out real network architectures with
abstract dynamic probes, as described in \cite{marr07_pre}, is a
promising approach in order to understand the link between topology
and dynamics. The question, if our dynamic probe is an appropriate
surrogate for a network's response to fluctuations, however
remains. It would thus be interesting to see the methodology of
\cite{holme07_pp} applied to the original MZ reaction sets, or more
appropriate network representations. In that respect, we agree with
the call of Holme and Huss for a more detailed type of modeling.

We thank Niko Sonnenschein (Darmstadt) for valuable discussions.

\end{document}